\documentstyle[prd,aps,preprint]{revtex}
\begin{document}
\draft

%
%

\preprint{Nisho-98/1} \title{Energy Dissipation of Axionic Boson Stars\\
 in Magnetized Conducting Media} 
\author{Aiichi Iwazaki}
\address{Department of Physics, Nishogakusha University, Shonan Ohi Chiba
  277,\ Japan.} \date{February 28, 1998} \maketitle
\begin{abstract}
Axions are possible candidates of dark matter 
in the present Universe. They have been argued to form
axionic boson stars with small masses 
$\sim 10^{-12}M_{\odot}$.
Since they possess oscillating electric fields 
in a magnetic field,
they dissipate their energies in magnetized
conducting media. We show that colliding with
a magnetized white dwarf, the axionic boson stars dissipate their energies  
and heat the white dwarf. Consequently the white dwarf cooled sufficiently
can emit detectable amount of radiations with the collision. 
Using a recent evaluation of the population of the white dwarfs
as candidates of MACHOs, we estimate     
the event rate of the collisions and obtain 
a result that the rate is large to be detectable.
\end{abstract}
\vskip2pc
The axion is the Goldstone boson associated with 
Peccei-Quinn symmetry\cite{PQ}, 
which was introduced to solve naturally the strong CP problem. 
In the early Universe some of the axions
condense and form topological objects\cite{kim,text}, i.e. 
strings and domain walls, 
although they decay below the temperature of QCD phase transition. 
After their decay, however, they have been 
shown to leave a magnetic field\cite{iwa} 
as well as cold axion gas as relics in 
the present Universe; the field is a candidate of a primordial
magnetic field leading to galactic magnetic fields observed 
in the present Universe.

In addition to these topological objects,  
the existence of axionic boson stars has been argued\cite{hogan,kolb}.
It have been shown numerically\cite{kolb} that in the early Universe, 
axion clumps are formed around the period of 
$1$ GeV owing to both the nonlinearity of an axion potential and 
the inhomogeneity of coherent axion oscillations
on the scale beyond the horizon. These clumps are called axitons 
since they are similar to solitons in 
a sense that its energy is localized. Then,
the axitons contract gravitationally to axion miniclusters\cite{kolb2}
after separating out from the cosmological expansion.
They are incoherent axions bound loosely, while 
the axion gases are distributed uniformly; they are 
generated by the decay of the axion strings 
or the coherent axion oscillations.
Furthermore, depending on energy densities, 
some of these miniclusters may contract gravitationally 
to coherent boson stars\cite{Tk,kolb,re}.
Their masses have been estimated roughly to be order of 
$\sim 10^{-12}M_{\odot}$. 
Eventually we expect that in the present Universe, 
there exist the axion miniclusters and the axion boson stars
as well as the incoherent axion gas as dark matter candidates. 
It has been estimated\cite{fem} that a fairly amount of the fraction of 
the axion dark matter is composed of the axion miniclusters and the axion 
boson stars. A way of the observation of 
the axion miniclusters has been discussed\cite{fem}.

In this letter we wish to point out an intriguing 
observable effect associated with 
the coherent axionic boson stars; we call them axion stars. 
Namely, they 
dissipate their energies in magnetized conducting media so that
the temperature of the media increases and strong radiations are expected. 
The phenomena are caused by electric fields generated by the  
coherent axion stars
under external magnetic fields. The electric fields induce electric currents
in the conducting media and loose their energies owing to the existence of 
resistances. Consequently the axion stars dissipate their energies in the 
magnetized conducting media.
Although the electric fields themselves are small,
the total amount of the energy dissipation is very large 
because the dissipation
arises all over the volume of the axion stars: Radii of the axion stars of 
our concern are such as $10^8\mbox{cm}\sim10^{10}\mbox{cm}$.
Consequently detectable amount of radiations are expected 
from the media heated in this way. 
Because the strength 
of the electric fields is proportional to the strength of the magnetic field,
the phenomena are revealed especially
in strongly magnetized media such as neutron stars, white dwarfs e.t.c..
We show that the amount of the energy dissipated in white dwarfs, 
for instance, with mass $\sim 0.5\,M_{\odot}$ and with 
magnetic field larger than $10^{5}$ Gauss is approximately given by 
$10^{35}$ erg/s $(M/10^{-14}M_{\odot})^4(m/10^{-5}\mbox{eV})^6$ 
where $M\, ( M_{\odot} )$ is the 
mass of the axion star ( the sun ) and $m$ is the mass of the axion.    
In our discussions we assume that the axion stars, when they collide with 
the media, are not destructed and that they pass the media. Later we 
discuss on this point.

Let us first sketch to review our solutions of the axionic boson stars.
Originally Seidel and Suen \cite{real} 
have found solutions of a real scalar axion field, $a$,
coupled with gravity. Their solutions represent 
spherical oscillating axion stars
with masses of the order of $10^{-5}M_{\odot}$; the solution $a$  
possesses oscillation modes with various frequencies.
On the other hand, axion stars of our concern are ones with much smaller 
masses, $\sim 10^{-12}M_{\odot}$.
So, in order to find the existence of such solutions
and explicit relations 
among the parameters, e.g. radius, $R$, mass, $M$, e.t.c., 
of these axion stars,  
we have numerically obtained solutions 
of the spherical axionic boson stars \cite{iwaza,real} in a 
limit of a weak gravitational field. Relevant equations are 
a free field equation
of the axion and Einstein equations. 
It means that our solutions represent
the axion stars with small masses, e.g. $10^{-12}M_{\odot}$; 
their gravitational fields are sufficiently weak and 
amplitudes $a$ are much small. Thus nonlinearity of the axion potential
is irrelevant; we have found that the nonlinearity arises  
only for the axion stars with masses larger than $\sim 10^{-9}M_{\odot}$.
The field $a$ of our solutions possesses only one oscillation mode 
with its frequency approximately given by $m$.
Other oscillation modes, 
which exist in general solutions representing the axion stars 
with larger masses, 
appear gradually as the masses increase. 
We have confirmed that our numerical solutions may be approximated by
the explicit formula,

\begin{equation}
\label{a}
a=f_{PQ}a_0\sin(mt)\exp(-r/R)\quad, 
\end{equation}
where $t$ ( $r$ ) is time ( radial ) coordinate and 
$f_{PQ}$ is the decay constant of the axion. 
The value of $f_{PQ}$ is constrained from cosmological 
and astrophysical considerations\cite{text} such as 
$10^{10}$GeV $< f <$ $10^{12}$GeV.

In the limit of the small mass of the axion star we 
have found a simple relation \cite{iwaza} between the mass, $M$ 
and the radius, $R$ 
of the axion star, 

\begin{equation}
\label{mass}
M=6.4\,\frac{m_{pl}^2}{m^2R}\quad,
\end{equation} 
with Planck mass $m_{pl}$.
Numerically, for example, 
$R=1.6\times10^5m_5^{-2}\mbox{cm}$ 
for $M=10^{-9}M_{\odot}$, 
$R=1.6\times10^8m_5^{-2}\mbox{cm}$ for $M=10^{-12}M_{\odot}$,
e.t.c. with $m_5\equiv m/10^{-5}\mbox{eV}$. 
A similar formula has been obtained in the case of boson stars of complex 
scalar fields.
We have also found an explicit relation \cite{iwaza} 
between the radius and the dimensionless amplitude $a_0$ in eq(\ref{a}),

\begin{equation}
\label{a_0}
a_0=1.73\times 10^{-8} \frac{(10^8\mbox{cm})^2}{R^2}\,
\frac{10^{-5}\mbox{eV}}{m}\quad.
\end{equation}
These explicit formulae are used 
for the evaluation
of the dissipation energy of the axion stars in the magnetized 
conducting media.

We now proceed to explain how the axion field representing these axion stars
generates an electric field
in an external magnetic field.
The point is that the axion couples with the electromagnetic fields
in the following way,

\begin{equation}
   L_{a\gamma\gamma}=c\alpha a\vec{E}\cdot\vec{B}/f_{PQ}\pi
\label{EB}
\end{equation}
with $\alpha=1/137$, where 
$\vec{E}$ and $\vec{B}$ are electric and magnetic fields respectively. 
The value of $c$ depends on the axion models\cite{DFSZ,hadron};
typically it is the order of one.

It follows from this interaction that Gauss law is 

\begin{equation}
\label{Gauss}
\vec{\partial}\vec{E}=-c\alpha \vec{\partial}(a\vec{B})/f_{PQ}\pi
+\mbox{``matter''}
\end{equation}
where the last term ``matter'' denotes contributions from ordinary matters.
The first term in the right hand side 
represents a contribution from the axion.
Thus it turns out that
the axion
field has an electric charge density, 
$-c\alpha\vec{\partial}a\cdot\vec{B}/f_{PQ}\pi$, 
under the magnetic field $\vec{B}$\cite{Si}. 
We assume that the field $\vec{B}=(0,0,B)$ is spatially uniform and 
that the field $a$ representing the axion stars 
is given by our solutions. Then we understand that 
this axion star has a charge distribution such that   
it has negative charges on a hemisphere 
( $z>0$ ) and positive charges on the other hemisphere ( $z<0$ ). 
Net charge is zero. Therefore the star possesses the electric field, 
$\vec{E}$, parallel to the magnetic field associated with 
the charge distribution; the electric field is given such that 
$\vec{E}=-c\alpha a\vec{B}/f_{PQ}\pi$.  
This field induces an electric current in conducting media and 
the energy of the field is dissipated.

Denoting the conductivity of the media by $\sigma$ and assuming 
the Ohm law, we find that
the axion star with it's radius $R$ dissipates an energy $W$ per unit time, 

\begin{eqnarray}
W&=&\sigma \alpha^2c^2B^2a_0^2R^3/\pi=
\alpha^2c^2B^2a_0^2R^3/4\pi^2\nu_m\\
&=&2.2\times 10^{21}\mbox{erg/s} 
\,\frac{c^2}{\nu_{m}/\mbox{cm$^2$s$^{-1}$}}\,\frac{M}{10^{-14}M_{\odot}}
\,\frac{B^2}{(1G)^2}\quad,
\label{W}
\end{eqnarray}
with magnetic diffusivity, $\nu_m=1/4\pi\sigma$.
We have used the explicit formulae eq(\ref{a}), eq(\ref{mass}) 
and eq(\ref{a_0}). 
Since the field $a$ 
oscillates\cite{iwaza,real} with a frequency given approximately by
the mass of the axion $m$, we have 
taken an average in time over the period, $m^{-1}$. 
Here, length scales of the media have been assumed to be 
larger than the radius 
of the axion star $R$; the whole of the axion star is included in the media. 
On the contrary, when the scale $L$ of the medium is smaller than $R$,
we need to put a volume factor of $(L/R)^3$ on $W$; only the fraction 
$(L/R)^3$ of the volume of the axion star is relevant for the dissipation.

We comment that the formula may be applied to the conducting media where the 
Ohmic law is hold even for oscillating electric fields with their frequencies
$m= 10^{10}\sim 10^{12}$ Hz. The law is hold in the media where 
electrons interact sufficiently many times in a period of $m^{-1}$ 
with each others or other charged 
particles and
diffuse their energies acquired from the electric field. 
Actually the law is 
hold in the convection zone of the sun, 
white dwarfs, neutron stars e.t.c..

We would like to point out that 
although the electric field 
$\vec{E}=-c\alpha a\vec{B}/f_{PQ}\pi$ is much small owing to 
the large factor of $f_{PQ}$,
the amount of the dissipation energy $W$ becomes large. The reason is that 
such a large value of the energy is resulted from 
the dissipation arising over the large volume; $W$ is proportional to 
$R^3$.

In order to evaluate the value of $W$, 
we need to know the mass of the axion star realized in 
the Universe. According to a creation mechanism of the axion star by 
Kolb and Tkachev\cite{kolb,kolb2,Tk}, 
the axionic boson stars are formed by some of miniclusters contracting
gravitationally. They have shown that the mass of the 
axion stars is typically $10^{-12}M_{\odot}\Omega_ah^2$; 
$\Omega_a$ is the ratio of the 
axion energy density to the critical density in the Universe and 
$h$ is Hubble constant in the unit of $100$ km s$^{-1}$ Mpc$^{-1}$.
Actual values of the mass
may range from $10^{-12}M_{\odot}$ to $10^{-14}M_{\odot}$ corresponding to 
the value, $0.01\le \Omega_ah^2\le 1$. 
Hence in this paper we discuss the axion stars 
with such a range of the masses.
Correspondingly the radius of the axion stars ranges 
from $10^8\mbox{cm}$ to $10^{10}\mbox{cm}$.

Now we discuss what amount of the axion energy is dissipated 
in a real magnetized conducting medium. As a first example we take the sun 
which is a typical star with 
a strong magnetic field $10^3\sim10^4$ G in its convection zone of the sun.
Assuming the depth of the convection zone being $\sim 2\times 10^{10}$ cm,
and the magnetic diffusivity $\nu_m\sim 10^7$ cm$^2$s$^{-1}$\cite{Zeld},
it follows that 

\begin{equation}
W=5.5\times 10^{22}\mbox{erg/s}\,
\frac{B^2}{(5\times 10^3\mbox{G})^2}\,\frac{M}{10^{-13}M_{\odot}}\quad,
\end{equation}
where we have set $c^2=1$. 
Hence the total energy dissipated, when the axion star passes the sun, 
is given such that

\begin{equation}
W_t=4\times 10^{10}\mbox{cm}\times W/v=
7.3\times10^{25}\mbox{erg}\,
\frac{B^2}{(5\times 10^3\mbox{G})^2}\,\frac{M}{10^{-13}M_{\odot}}\quad,
\end{equation}
where the velocity, $v$, of the axion star is assumed to be 
$3\times 10^7\mbox{cm/s}$ ; this is a value obtained by equating
a kinetic energy with a gravitational energy of the axion star in our galaxy.
We have assumed the absence of the magnetic field 
in the radiation zone. Therefore, the effect of the energy dissipation  
of the axion star in the sun is difficult to be observed;
the luminosity of the sun is the order of $10^{33}$ erg/s. 
Similarly, as far as we are concerned with 
such stars with the same physical parameters as those of the sun, 
it is difficult to detect the effect of the energy dissipation.


Next we go on to discuss the case of white dwarfs which may possess
strong magnetic field $\sim 10^6$ G with their typical radius 
$L\sim10^9$ cm. The radius is larger than that ( $R\sim 10^8$ cm )
of the axion star
with mass $\sim 10^{-12}M_{\odot}$, but smaller than that 
( $R\sim 10^{10}$ cm ) of the axion 
star with mass $\sim 10^{-14}M_{\odot}$. 
So we need to treat separately the two cases.
But since 
the recent observations indicate much smaller values of $\Omega h^2$ than 
$1$ ( $\Omega$ is the ratio of the energy density to the critical density ),
here we consider the case of the axion star with mass 
$\sim 10^{-14}M_{\odot}$, which corresponds to 
the case, $\Omega_ah^2\sim 0.01$.
Thus the radius of the axion star is 10 times larger than 
that of the white dwarf.
( It will be turned out that with such a choice of 
small masses $\sim 10^{-14}M_{\odot}$,
the rate of the white dwarfs colliding with 
the axion stars is large to be observed in our galaxy.)

According to recent observations of gravitational
microlensing\cite{macho}, the white dwarfs are most plausible candidates of 
the phenomena. Most of them seem to be dark enough not to be 
observed. Probably, they have reached a stage of a fast Debye cooling
and have lost almost of all their internal thermal energies\cite{WD}.
Although this conjecture might be not correct,
we assume it with an additional assumption that the dark white 
dwarfs have strong magnetic field $\sim 10^6$ G.

Then, the observation indicates that 
the population of the white dwarfs 
is $2\times10^{11}M_{\odot}/0.5\,M_{\odot}\sim 4\times10^{11}$
in the halo of our galaxy when their typical mass is
$0.5\,M_{\odot}$.     
Since the number is so large, the collision with the axion stars
is expected to occur frequently in our galaxy. 
Furthermore, as we will show soon,
the axion star dissipates the energy 
in the white dwarf so much
that the effect of the dissipation in the white dwarfs with very low
temperature
may be observed in a search like  
MACHO searches.

When we apply naively the formula eq(\ref{W}),
the amount of the energy dissipated in the white dwarf is 
given such that

\begin{equation}
\label{w}
W\sim 10^{30}\mbox{erg/s} 
\,\frac{c^2}{\nu_{m}/\mbox{cm$^2$s$^{-1}$}}\,\frac{M^4}{(10^{-14}M_{\odot})^4}
\,\frac{B^2}{(10^6\mbox{G})^2}\,\frac{m^6}{(10^{-5}\mbox{eV})^6}\quad.
\end{equation}
Here we have taken account of the volume factor, 
$(L/R)^3\sim$ $10^{-3}(M/10^{-14}M_{\odot})^3m_5^6$, 
because the radius $R(\sim 10^{10}\mbox{cm})$ of the 
axion star is larger than the radius $L(\sim 10^9\mbox{cm})$ 
of the white dwarf. 
We expect that $\nu_m$
is much smaller ( or conductivity is much larger ) than 
ones of normal metals because 
the number density of degenerate electrons in the white dwarf 
is much larger than that of the metals.
According to theoretical evaluations\cite{con} 
it follows that $\nu_m\sim O(10^{-5})$cm$^2$/s in the case of 
a crystallized white dwarfs with temperature $\sim10^4$K 
and density $10^6$ g/cm$^3$. Note that lower temperature leads to 
smaller $\nu_m$ ( larger conductivity ).  
Since our concern is dark white dwarfs 
with very low temperature as we mentioned before, 
actual values of $\nu_m$ are expected to be much lower than 
the above one.

Then $W$ reaches a value more than $10^{35}$erg/s.
But the dissipation of such a large amount of the energy more than  
$W\sim10^{35}$erg/s, is impossible 
owing to the energy conservation. Namely, the maximum energy, which 
a part of the axion star swept by the white dwarf 
can dissipate, is the energy stored in the part;
we note that the white dwarf is 
smaller than the axion star. 
We find that the energy stored in the part is given by 
$3L^2vM/4R^3\sim 10^{35}\mbox{erg/s}\,(M/10^{-14}M_{\odot})^4m_5^6$.
This is smaller than $W$ estimated naively.
Thus real amount of the energy dissipated is at most
given by
$W_{real}\sim10^{35}$ erg/s $(M/10^{-14}M_{\odot})^4m_5^6$. 
The dissipation of the energy continues 
until the white dwarf passes the axion star. 
Thus total energy dissipated by the axion star is 
$2 R/v\times W_{real}\sim 10^{38}\mbox{erg}\,(M/10^{-14}M_{\odot})^3m_5^4$.


Since we suppose that the white dwarfs of our concern have lost most of 
their internal energies by cooling and that their temperature, $T$, is 
sufficiently low, their specific heat, $c_v$, per ion
is given approximately by \cite{c_v} 
$c_v\sim 16\pi^4(T/\theta_D)^3/5$,
where $\theta_D$ is the Debye temperature, typically being $10^7$ K.
Hence the injection of the energy, 
$10^{38}\mbox{erg}\,(M/10^{-14}M_{\odot})^3m_5^4$, 
increases the temperature of the white dwarfs to 
$\sim 10^4\mbox{K}\,(M/10^{-14}M_{\odot})^{3/4}\,m_5$. 
If surface temperature is the same as 
this temperature, the luminosity is roughly 
$10^{-3}L_{\odot}\,(M/10^{-14}M_{\odot})^3m_5^4$. 
Although this estimation is too naive, the result is interesting because 
the luminosity
is large enough to be observed.

To analyze more precisely the phenomena, however, 
we need to solve several questions. 
Among them, we have to make clear whether or not a gravitational 
attraction of the white dwarfs destabilizes the axion stars, which 
are composites of axions bound gravitationally.
We expect that since the axion stars are coherent objects, their stability 
is not affected by the attraction, although the original spherical 
distribution of the field $a$ is fairly distorted.
Namely the coherent axions will not be transformed into incoherent 
axions by the slowly changing gravitational perturbations. 
Therefore even if we include the gravitational effects of the white dwarfs,
similar amount of the dissipation energy is expected to be released 
from the axion star  
although the geometrical form of the star is fairly deformed.
More difficult problem is to analyze the back reaction of 
the energy dissipation on the axion stars and to ask the question 
whether or not they are destabilized or trapped by the white dwarfs 
with the effects of the dissipation. It seems that both effects 
of the gravitational attraction and of the dissipation make 
the axion stars be trapped owing to the dissipation of their kinetic 
energies. In this case more energies are released from 
the axion stars, compared with the case of they passing the white dwarfs.
Consequently, stronger radiations are expected. 
The analysis of these problems will be presented in future publication.

We have shown that the collision between the axion star and the invisible 
white dwarf makes the white dwarf become visible. Thus it is important 
to see how large the rate of the collision is. 
We estimate the rate of such a collision in our galaxy. 
Especially, we are concerned 
with the event rate observed
in a solid angle,
$5^{\circ}\times 5^{\circ}$, for example. We assume that 
as indicated by the recent observations of gravitational
microlensing, the half of the halo
is composed of the white dwarfs with mass 
$0.5\times M_{\odot}$. The other
half is assumed to be composed of the axion stars. 
Total mass of the halo is supposed
to be $\sim 4\times 10^{11}M_{\odot}$. 
Furthermore, the distribution\cite{dis} of the halo 
is taken such that its density 
$\propto (r^2+ 3 R_c^2)/(r^2+R_c^2)^2$ with $R_c=4$ kpc 
where $r$ denotes a radial coordinate with the origin being the center of 
the galaxy ( the final result does not depend practically 
on the value of $R_c=2\sim 8$ kpc ).
Then it is easy to evaluate the event rate of the collisions,

\begin{equation}
0.5/\mbox{year}\times\frac{(10^{-14}M_{\odot})^3}{M^3}\,
\frac{(10^{-5}\mbox{eV})^4}{m^4}\,\frac{\Omega}{5^{\circ}\times5^{\circ}}
\label{rate}
\end{equation}
where $\Omega$ is a solid angle.  
We have taken into account the fact that the earth is located at about 
$8$ kpc from the center of our galaxy, simply by 
counting the number 
of the collisions arising in the region
from $8$ kpc to $50$ kpc. Therefore, 
it is possible to observe the phenomena associated with 
the energy dissipation of the axion stars in the white dwarfs,
although the rate depends heavily on both of the mass 
of the axion stars and the axion mass.

Similarly as the white dwarf, neutron stars possess strong magnetic fields
and high electric conductivities. Thus the energy dissipation of the 
axion star is expected to be large in the neutron stars. But since 
the radius of the neutron star is about $\sim 10^6$ cm, the actual amount of 
the dissipation energy is much smaller than that in the white dwarf.
The reason is that the maximum energy, which can be dissipated, 
is the energy stored in the part of the axion star 
swept by the neutron star and that the volume of the part is smaller 
than that in the case of the white dwarf.
Therefore, 
the rate of the energy dissipated is approximately 
given by $10^{35}(10^6/10^9)^2$ erg/s $(M/10^{-14}M_{\odot})^4m_5^6$ 
$\sim10^{29}$ erg/s $(M/10^{-14}M_{\odot})^4m_5^6$. 
Total amount of the energy is $10^{32}$ erg $(M/10^{-14}M_{\odot})^3m_5^4$.
This is too small for the resultant radiation to be detectable.

As explained in above examples, the axion stars are possible sources
for generating energies in the magnetized conducting media. We may apply 
the idea to systems such as accretion disks with strong magnetic fields 
around black holes e.t.c.. Probably the existence of the axion
will be confirmed indirectly by observing the phenomena associated with
the energy dissipation of the axion star.

In summary, we have shown that the coherent axion stars dissipate their 
energies in the magnetized conducting media such as the sun,
or the white dwarfs. Among them, the white dwarfs with much low 
temperature are heated in this mechanism and emit the radiations with
detectable luminosities. Furthermore, we have shown that 
the rate of the white dwarfs
colliding with the axion stars is not small in our galaxy. 
Therefore the collision may be detected in a rearch of the phenomena 
like a noba. Although the possibility of these phenomena being observable 
depends sensitively on the physical parameters such as the axion mass, 
the mass of the axion star,
or the strength of the magnetic field, it is interesting that 
there is an allowed range of the parameters making 
the phenomena detectable.

Principal part of this work have been down when the author has visited 
the Theoretical Physics Group at LBNL. He would 
like to express his thank for useful discussions and comments to 
Professors M. Kawasaki and R. Nishi, and also for
the hospitality in LBNL as well as in  
Tanashi KEK.





\end{document}